\documentclass[preprint,12pt,3p]{elsarticle}

\usepackage{amssymb}
\usepackage{amsmath}
\usepackage{mathrsfs}
\usepackage[draft]{todonotes}

 \usepackage{lineno}

\journal{Astroparticle Physics}

\begin{document}
\begin{frontmatter}

\title{Calibration of the LOFAR low-band antennas using the Galaxy and a model of the signal chain}

\address[label1]{ Astrophysical Institute, Vrije Universiteit Brussel, Pleinlaan 2, 1050 Brussels, Belgium}
\address[label2]{Department of Astrophysics/IMAPP, Radboud University, P.O. Box 9010, 6500 GL Nijmegen, The Netherlands}
\address[label3]{Nikhef, Science Park 105, 1098 XG Amsterdam, The Netherlands}
\address[label4]{Netherlands Institute of Radio Astronomy (ASTRON), Postbus 2, 7990 AA Dwingeloo, The Netherlands}
\address[label5]{KVI-CART, University Groningen, P.O. Box 72, 9700 AB Groningen}
\address[label7]{Department of Astrophysical Sciences, Princeton University, Princeton, NJ 08544, USA}
\address[label8]{Interuniversity Institute for High-Energy, Vrije Universiteit Brussel, Pleinlaan 2, 1050 Brussels, Belgium}
\address[label9]{Department of Physics and Electrical Engineering, Linn\'euniversitetet, 35195 V\"axj\"o, Sweden}
\address[label10]{DESY, Platanenallee 6, 15738 Zeuthen, Germany}
\address[label11]{Institut f\"{u}r Kernphysik, Karlsruhe Institute of Technology(KIT), P.O. Box 3640, 76021, Karlsruhe, Germany}
\address[label12]{Institut f\"{u}r Physik,Humboldt-Universit\"{a}t zu Berlin, 12489 Berlin, Germany}

\author[label1]{K.~Mulrey\corref{cor1}}
\ead{kmulrey@vub.be}

\author[label2]{A.~Bonardi}
\author[label1,label2]{S.~Buitink}
\author[label2]{A.~Corstanje}
\author[label2,label3,label4]{H.~Falcke}
\author[label5]{B.~M.~Hare}
\author[label2,label3,label8]{J.~R.~H\"{o}randel}
\author[label1,label11]{T. Huege}
\author[label1]{P.~Mitra}
\author[label10,label12]{A.~Nelles}
\author[label2]{J.~P.~Rachen}
\author[label2]{L.~Rossetto}
\author[label2,label7]{P.~Schellart}
\author[label5,label8]{O.~Scholten}
\author[label2,label4]{S.~ter~Veen}
\author[label2,label9]{S.~Thoudam}
\author[label5]{T.~N.~G.~Trinh}
\author[label1]{T.~Winchen}

\cortext[cor1]{Corresponding author}




\begin{abstract}
The LOw-Frequency ARray (LOFAR) is used to make precise measurements of radio emission from extensive air showers, yielding information about the primary cosmic ray.  Interpreting the measured data requires an absolute and frequency-dependent calibration of the LOFAR system response.  This is particularly important for spectral analyses, because the shape of the detected signal holds information about the shower development.  We revisit the calibration of the LOFAR antennas in the range of $30-80$ MHz. Using the Galactic emission and a detailed model of the LOFAR signal chain, we find an improved calibration that provides an absolute energy scale and allows for the study of frequency dependent features in measured signals.  With the new calibration, systematic uncertainties of $13\%$ are reached, and comparisons of the spectral shape of calibrated data with simulations show promising agreement.
\end{abstract}


\end{frontmatter}

\section{Introduction}\label{sec:intro}
Radio emission from extensive air showers has proven to be an effective way to measure energetic cosmic rays.   Radio signals carry information about the development of the air shower, and are used to reconstruct shower properties such as the energy and atmospheric depth of the shower maximum ($X_{\textrm{max}}$)~\cite{Apel:2014jol,buitink2014,Aab:2015vta,Bezyazeekov:2015ica}.  Features of the radio emission are now well understood~\cite{huege2016}.  The LOw-Frequency ARray (LOFAR) has been successfully used to detect air showers since 2011~\cite{LOFAR, schellart2013}.  Due to its dense antenna spacing, LOFAR is able to sample the radio emission in great detail.  Precise measurements have been made of $X_{\textrm{max}}~$\cite{buitinkNature2016}, wavefront shape~\cite{Corstanje:2014waa}, and the circular polarization in the shower~\cite{Scholten:2016gmj}.  There are ongoing efforts to understand the shape of the power spectra of radio emission~\cite{rossettoICRC2017}.  This observable is sensitive to the development of the air shower and is another way to glean information about $X_{\textrm{max}}$ and shower development~\cite{jansenThesis}.  

Interpreting LOFAR data requires knowledge of the electric field at the antenna.  The measured signal is compared to predictions from first-principle simulation packages which calculate radio emission from the electromagnetic component of air showers.  This component is well-understood and simulations have very small systematic uncertainties~\cite{Gottowik:2017wio}, so the uncertainties of the measured signal dominate in the analysis. For this reason, it is necessary to have a good understanding of the detector response, including gains and losses in the LOFAR signal chain, the directional response of the antennas, and dispersion in the system.  Knowledge of the directional response and dispersion in the antennas comes from simulations like WIPL-D~\cite{wiplD} and in-situ measurements where the directionality of the antenna can be explicitly measured.  In this work, we derive an absolute calibration that relates the recorded signal in analog-digital conversion (ADC) units to the voltage received at the antenna, correcting for frequency dependent gains and losses in different components of the signal chain.  Previously, two methods have been used to derive an absolute calibration of the low-band antennas (LBAs) which cover a frequency range of $30-80$ MHz~\cite{Nelles:2015gca}.

The first method makes use of a VSQ 1000 external reference source~\cite{teseq}.   The same reference source was used by the LOPES~\cite{Schroder:2010zz} and \mbox{Tunka-Rex}~\cite{Schroder:2017nkf} experiments, allowing measurements from different experiments to be compared.   The Pierre Auger Observatory used a similar external reference source attached to an octocopter, to achieve a calibration with an overall systematic uncertainty of 9\% \cite{Aab:2017lby}.  The drawback of using reference sources is that this method relies on the calibration of the reference source itself.  Calibrating antennas at low frequencies is difficult; the long wavelengths involved make it hard to calibrate the antenna in the far field.  The reference source was calibrated by the manufacturer between $30-80$ MHz in both a Standard Anechoic Chamber and a Gigahertz Transverse Electromagnetic (GTEM) cell~\cite{comm2}.  Neither calibration was performed in the far field of the antenna.  The two methods result in two different calibrations of the reference source.  Although the total power in the signal is similar between the methods, there are frequency dependent differences across the band, most notably between $55-75$ MHz (see \ref{sec:ref_source}).  These differences propagate into the LOFAR antenna calibration and affect the results of the offline data reduction, limiting the potential of frequency spectrum analyses.


The second method uses radio emission from the Galaxy as a calibration source.  Galactic emission is the dominant external background signal in the $30-80$ MHz band.  This emission, combined with thermal noise from the LOFAR electronics, is what makes up the background noise in the LOFAR cosmic-ray data, once the data have been cleaned for narrowband radio-frequency interference.  Electronic noise contributions play a critical role in the frequency dependence of the calibration.  Since data are collected regularly, there is a large sample of background data with which to calibrate the LBAs.  This method does not require a dedicated campaign and can be repeated anytime.  However, it relies on knowledge of the electronic noise in the LOFAR  signal chain and a precise knowledge of the Galactic emission as a function of local sidereal time (LST).  Without using a model of the LOFAR signal chain, the electronic noise could only be determined to within a systematic uncertainty of 37\%, which was prohibitively high~\cite{Nelles:2015gca}.

Both calibration methods give signal amplitudes that agree within systematic uncertainties.  However, there is a need for smaller uncertainties on the amplitude and a more precise understanding of the spectral shape of detected pulses.  In this work, the calibration using Galactic emission is revisited.  We characterize the electronic noise in detail, yielding an absolute, frequency dependent calibration of the LBAs with reduced systematic uncertainties. 

This paper is organized as follows.  Section~\ref{sec:lofar} introduces the LOFAR telescope and cosmic-ray data processing techniques.  Section~\ref{sec:calibration} covers the calibration method, including details of Galactic emission the LOFAR signal chain.  The instrumental noise in the signal chain, resulting calibration, and systematic uncertainties are presented in  Section~\ref{sec:results}.  Section~\ref{sec:discussion} includes a comparison of calibrated LOFAR data and Monte Carlo simulations and a discussion of the systematic effects seen in Tunka-Rex data and LOFAR calibrations.

\section{LOFAR}\label{sec:lofar}

LOFAR antenna stations are located across northern Europe with a dense core in the North of the Netherlands consisting of 24 stations~\cite{LOFAR}.  Six stations are located within an area of 160~m radius called the Superterp.  Each station includes 48 high-band antennas (HBAs) covering a frequency range of $110-240$ MHz and 96 low-band antennas (LBAs) which cover a range from $30-80$ MHz.  The LBAs are further divided into `inner' and `outer' configurations.  Each antenna is digitized at a rate of 200 mega-samples per second and the data are written to a circular  ring buffer called a transient buffer board (TBB).  For the purpose of detecting cosmic rays, a particle detector array of 20 scintillators was installed on the Superterp~\cite{thoudam2014,thoudam2016}.  When a cosmic ray is detected by the scintillator array, the TBBs are read out and the data are saved.  At a given time, either the HBAs, `inner' LBAs, or `outer' LBAs can be operational.  The focus of this work is the calibration of the LBAs.

Each LBA consists of two orthogonal inverted dipoles, X and Y, attached to a low-noise amplifier (LNA).  The signal of each dipole propagates through a coaxial cable to the receiver unit (RCU), where it is filtered, amplified and digitized.  The LOFAR signal chain is discussed in further detail in Section \ref{sec:cal_elec}.  
When a cosmic ray is detected, data from the operating LBAs are processed in a reconstruction pipeline, details of which are found in \cite{schellart2013}.  Two steps of particular interest to this work are the amplitude calibration and the antenna pattern unfolding.  The amplitude calibration provides units of watts for the recorded voltage trace, and corrects for frequency dependent components of the signal chain.  In order to relate the output voltage of the antenna to the physical electric field at the location of the antenna, the frequency and directionally dependent antenna response, or vector effective height, $\mathbf{H}(\nu,\theta,\phi)$,  must be unfolded from the signal.  For the LBAs, $\mathbf{H}(\nu,\theta,\phi)$ was simulated using WIPL-D software and includes amplification from the LNA~\cite{LOFAR,wiplD}.   It can be expressed using a Jones Matrix, which is discussed in \ref{sec:cal_theory}.  The amplitude calibration is the subject of this work.

\section{Calibration Technique}\label{sec:calibration}

The calibration factor $C(\nu)$ relates the recorded signal in ADC units to the expected signal output of the antenna in Volts.  It encompasses the absolute conversion factor between units, as well as the frequency dependent corrections that need to be made for gains and losses in the LOFAR signal chain.\footnote{Phase factors introduced in the LOFAR signal chain are not considered in this work.}  In terms of power, this can be expressed as
\begin{equation}\label{cal:eq}
    C^2(\nu)=\frac{P_e(\nu)}{P_m(\nu)}
\end{equation}
where  $P_m(\nu)$ is the power of measured LOFAR data and $P_e(\nu)$ is the expected power delivered to the load of the antenna.  In order to derive the calibration factor $C(\nu)$, a known source must be used.  Background traces from the LBAs are known to be dominated by sky noise, with contributions from thermal noise from the LOFAR electronics~\cite{LOFAR}.  Cosmic-ray data have been collected since 2011, and since the cosmic-ray signal is only a small fraction of the time traces, the remaining background provides the $P_m(\nu)$ necessary for the calibration.  This is discussed in Section \ref{sec:lofar_data}. 

If Galactic emission were the only contribution to the LOFAR background, the calibration factor would simply be the ratio of the power expected from the Galaxy to the recorded power.  However, electronic noise enters the system at various stages of the LOFAR signal chain.  The frequency content of background signal depends on the strength of the noise as well as where it enters the system, and so the noise must be well understood to produce the correct calibration.  The noise levels are not known exactly, and therefore a fitting procedure is used to find them, which makes use of the time variation of the Galactic emission visible at the LOFAR site.  This procedure is similar to the one originally used to find a Galactic calibration of the LOFAR antennas~\cite{Nelles:2015gca}.  The difference is in how the electronic noise is modeled.  The Galactic emission and modeling of the signal chain are discussed in Sections~\ref{sec:cal_gal} and \ref{sec:cal_elec}.

\subsection{LOFAR data}\label{sec:lofar_data}

LOFAR background traces are used to generate $P_m(\nu)$ in equation \ref{cal:eq}.  To date, data from more than 6000 cosmic-ray events have been recorded.  Most events are too low in energy to provide a usable cosmic-ray signal, however, the background can still be used for calibration purposes.  For this analysis, data from the six Superterp stations are used.  Only data from the `outer' configurations are used, since the antenna spacing is such that mutual coupling between antennas is not a concern.

For each recorded event, 2.1~ms of data are saved for each of the two dipoles of each antenna, which are treated separately.  Narrowband RFI is removed, and data are separated into blocks of 1024 samples~\cite{schellart2013}.  The time of the cosmic-ray signal is estimated using information from the particle detectors, and data within $\pm 5$ blocks of the expected cosmic ray signal are removed.  The Fourier transforms of 100 blocks are calculated and averaged for each event, providing 0.2~MHz resolution.   The average power in the background for each event is calculated as
\begin{equation}
    P_{m}(\nu)=\frac{|\mathscr{F}(\nu)|^2}{R_r},
\end{equation}
where $\mathscr{F}(\nu)$ denotes the Fourier transform of the measured signal in ADC units and $R_r$ is the radiative resistance of the antenna.  Events with anomalous power levels are excluded.  Because the Galactic emission is time-varying, the set of background data is divided into 15 minute intervals in local sidereal time (LST).  Data are also grouped into 1 MHz bins, in order to achieve a frequency dependent calibration.  On average, there are 7250 measured samples for each bin, from the 48 antennas of each of the Superterp stations.

\subsection{Galactic emission}\label{sec:cal_gal}

LFmap software is used to predict the radio emission coming from the Galaxy~\cite{LFmap}.  Existing sky maps at selected frequencies and equatorial coordinates are interpolated to the desired frequencies, making use of the fact that the brightness temperature of the sky follows a power law.  The temperature at a given frequency $\nu$, right ascension $\alpha$, and declination $\delta$, is modeled as
\begin{equation}
    T_{\mathrm{sky}}(\nu,\alpha,\delta)=T_{\mathrm{CMB}}+T_{\mathrm{Iso}}(\nu)+T_{\mathrm{Gal}}(\nu,\alpha,\delta).
\end{equation}
$T_{\mathrm{CMB}}$ is the temperature of the cosmic microwave background, and $T_{\mathrm{Iso}}(\nu)$ is the isotropic component of the temperature, which is thought to be due to integrated emission over unresolved sources.  $T_{\mathrm{Gal}}(\nu,\alpha,\delta)$ is the radio emission from known galactic sources.  In the frequency band $30-80$ MHz,  $T_{\mathrm{Gal}}(\nu,\alpha,\delta)$ is the dominating contribution.  Systematic uncertainties on the brightness temperature are discussed in Section~\ref{sec:results} and~\ref{sec:gal_modeling}.

The spectral radiance at the antenna is described by the Rayleigh-Jeans Law,
\begin{equation}\label{eq:rayleigh_jean}
B(\alpha,\delta,\nu)=\frac{2k_B}{c^2}\nu^2T_{\mathrm{sky}}(\alpha,\delta,\nu)
\end{equation}
where the spectral density is the integral over the visible solid angle
\begin{equation}\label{eq:spectral_power_density}
S({\nu})=\int_{\Omega}B(\alpha,\delta,\nu)d\Omega=\frac{2k_B}{c^2}\nu^2\int_{\Omega}T_{\mathrm{sky}}(\alpha,\delta,\nu)d\Omega. 
\end{equation}
Before integrating over the visible sky, the $(\alpha,\delta)$ equatorial coordinates are converted to local $(\theta,\phi)$ celestial coordinates as defined in \ref{sec:cal_theory}.  This introduces a time dependence.  The power delivered to the load of the antenna at a given frequency and time is
\begin{equation}\label{eq:powerTA}
\begin{split}
	P_{\mathrm{sky}}(t,\nu)&=\frac{2k_B}{c^2}\int_{\nu}\nu^2 \int_{\Omega} T_{\mathrm{sky}}(t,\nu,\theta,\phi)A_e(\nu,\theta,\phi)  d\nu d\Omega\\
	&=\frac{2k_B}{c^2}\int_{\nu}\nu^2 \int_{\Omega} T_{\mathrm{sky}}(t,\nu,\theta,\phi)\frac{|\mathbf{H}(\nu,\theta,\phi)|^2 Z_0}{R_r}  d\nu d\Omega
\end{split}	
\end{equation}
where $A_e(\nu,\theta,\phi)$ is the effective area of the antenna, and is related to $\mathbf{H}(\nu,\theta,\phi)$ as described in \ref{sec:cal_theory}.  The average antenna response to unpolarized waves, for dipole X of the LBA, is

\begin{equation}
	\langle|\mathbf{H}_X(\nu,\theta,\phi)|^2\rangle=\frac{1}{2}(|J_{X\theta}(\nu)|^2+|J_{X\phi}(\nu)|^2) 
\end{equation}
where $J_{X\theta}(\nu)$ and $J_{X\phi}(\nu)$ are components of the Jones matrix describing the LBA response~\cite{schellart2013}.   This can be written as an antenna gain term, $G_{\mathrm{ant}}(\nu)$, and a directional dependence term, $D(\theta,\phi)$, so that 
\begin{equation}
	\langle|\mathbf{H}_X(\nu,\theta,\phi)|^2\rangle=\frac{1}{2}G_{ant}(\nu)D(\theta,\phi). 
\end{equation}
Here the X dipole was used as an example.  The $J_{Y\theta}(\nu)$ and $J_{Y\phi}(\nu)$ components of the Jones matrix are used in the case of the Y dipole.  The full form of the power in one dipole is  
\begin{equation}\label{total_sky_power_eq}
P_{\mathrm{sky}}(t,\nu)=\frac{k_BZ_0}{c^2R_r}\int_{\nu}\nu^2\int_{\Omega} T_{\mathrm{sky}}(t,\nu,\theta,\phi)G_{\mathrm{ant}}(\nu)D(\theta,\phi)d\nu d\Omega. 
\end{equation}
For the purposes of adding noise values to the Galactic signal, in the following sections $P_{\mathrm{sky}}(t,\nu)$ only includes the directional component of the antenna model, $D(\theta,\phi)$.  The gain term $G_{\mathrm{ant}}(\nu)$ is introduced in the modeling of the signal chain. Uncertainties on the antenna pattern are discussed in Section \ref{sec:results}.

\subsection{Contribution of Signal Chain Components}\label{sec:cal_elec}

\begin{figure}[h!]
\centering
\includegraphics[scale=0.45,trim={3cm 0cm 3cm 0},angle =90]{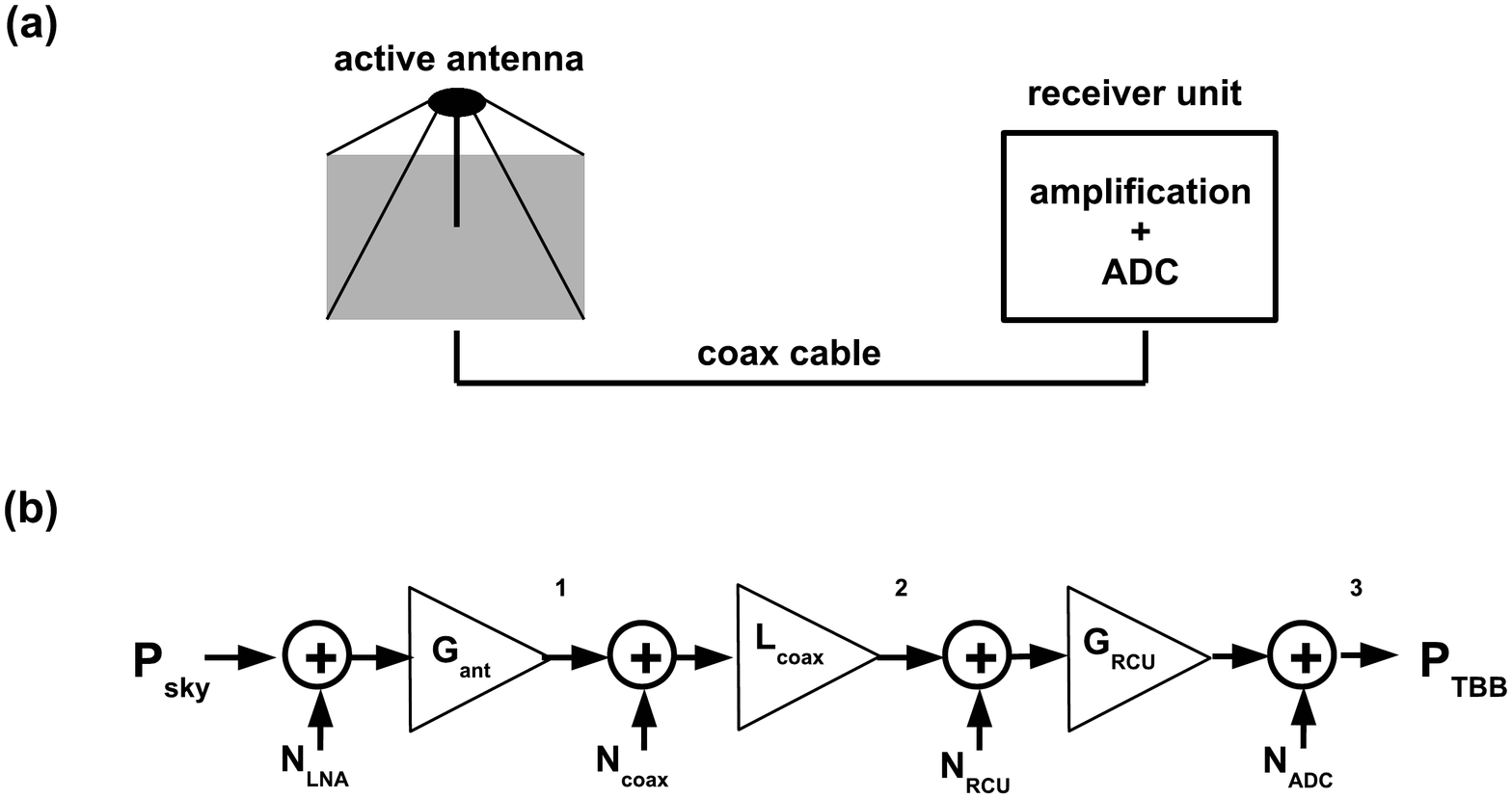}
\caption{LOFAR signal chain, following sky power, $P_{\mathrm{sky}}$, to power recorded from the TBBs, $P_{\mathrm{m}}$.  (a)~Schematic view of the three primary parts of the signal chain.  (b)~Detailed view of gains, losses, and electronic noise additions.  Each contribution is discussed and the numbers 1, 2, 3 are referenced by the equations in Section \ref{sec:cal_elec}.  Note that the directionality of the antenna is already folded into $P_{\mathrm{sky}}$, so that the noise from the LNA, $N_\mathrm{LNA}$, only propagates through the amplification from the LNA and doesn't include a directional component.}
\label{lba_signal_chain}
\end{figure}

The sky power is propagated through the signal chain, adding electronic noise where needed.  The result of this modeling yields the simulated power, $P_{\mathrm{sim}}(t,\nu)$ which can be directly compared to $P_{\mathrm{m}}(t,\nu)$. Electronic noise is expected to have a flat frequency spectrum to first order.  However, noise contributions from different parts of the signal chain propagate through frequency dependent losses and gains, and so the final noise contribution is indeed frequency dependent.  In order to model the noise, each step of the signal chain is considered separately.  There are three main contributions, as shown in the top of Figure~\ref{lba_signal_chain}.  The signal propagates through an active antenna, through coaxial cable, and then into the RCU, where it is amplified and digitized.

Individual gains and losses, as well as the electronic noise injected into the system, are shown in the bottom of Figure~\ref{lba_signal_chain}. The following subsections detail components in the signal chain and build an expression for the simulated background signal.

\subsubsection{Active antenna}

The first step of the signal chain is the active antenna.  The antenna response, including both directionality, $D(\theta,\phi)$, and gain, $G_{\mathrm{ant}}(\nu)$, was simulated using a WIPL-D simulation.  We start with $P_{\mathrm{sky}}(t,\nu)$, which already contains the directional response of the antenna.  Noise generated in the active antenna, $N_{\textrm{LNA}}$, is added to the predicted sky noise.   Both are then multiplied by the antenna gain, $G_{\mathrm{ant}}(\nu)$.  We add the noise at this point so that it propagates through the amplifier of the antenna, but does not contain directional information.  A frequency dependent correction factor, $A(\nu)$, is added at this stage to correct for adjustments needed in the antenna model, including a shift in resonance frequency.  The power at the active antenna can then be written as 

\begin{equation}
    P_{\mathrm{sim,1}}(t,\nu)=\bigg( P_{\mathrm{sky}}(t,\nu) + N_{\mathrm{LNA}}\bigg) G_{\mathrm{ant}}(\nu) A(\nu).
\end{equation}

The subscripts $P_{\mathrm{sim,1}}$, $P_{\mathrm{sim,2}}$, $P_{\mathrm{sim,3}}$ in this and the following equations reference the points in the signal chain where the power is calculated, as indicated in Figure~\ref{lba_signal_chain}.  The left panel of Figure~\ref{fig:cable_attenna} shows the antenna gain as simulated with WIPL-D software.  Here, the resonance frequency is visible close to 58 MHz.   The correction $A(\nu)$ is not a simulated or measured quantity, but is found as a result of the fitting procedure.

\subsubsection{Coaxial cable}

\begin{figure}[h!]
\centering
\includegraphics[scale=0.35]{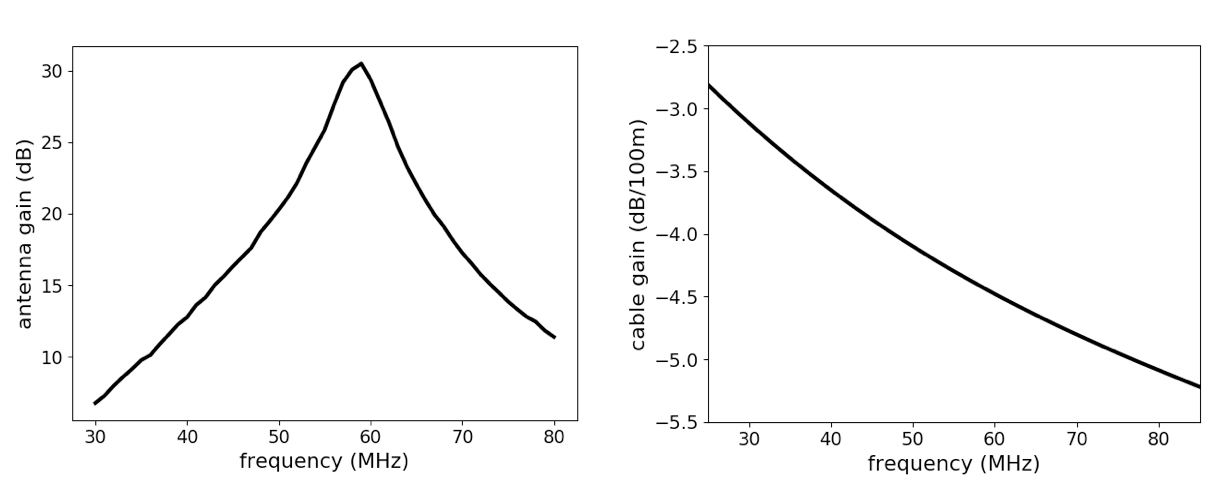}
\caption{Left: Gain of the LBAs, $G_{\mathrm{ant}}(\nu)$, simulated with WIPL-D software~\cite{wiplD}. Right: Gain for a 100~m coaxial cable, $L_{\mathrm{coax}}(\nu)$~\cite{coax9}.}
\label{fig:cable_attenna}
\end{figure}

The signal travels from the antenna to an electronics cabinet located at each LOFAR station through coaxial cable, which has a frequency dependent attenuation factor obtained from manufacturer specifications, as seen in the right panel of Figure~\ref{fig:cable_attenna}~\cite{coax9}.  Cables are of different lengths, which must be taken into account for each antenna.  There is expected to be a noise contribution, $N_{\mathrm{coax}}$,  in the cables.  However, based on LOFAR estimates~\cite{LOFAR_receiver_system}, it is small compared to other contributions, and so is not included in the model ($N_{\mathrm{coax}}=0$).  A frequency dependent loss term is added to the simulated noise,

\begin{equation}
    P_{\mathrm{sim,2}}(t,\nu)=\bigg( P_{\mathrm{sky}}(t,\nu) + N_{\mathrm{LNA}}\bigg) G_{\mathrm{ant}}(\nu) A(\nu)L_{\mathrm{coax}}(\nu).
\end{equation}

\subsubsection{Receiver Unit}

\begin{figure}[h!]
\centering
\includegraphics[scale=0.48]{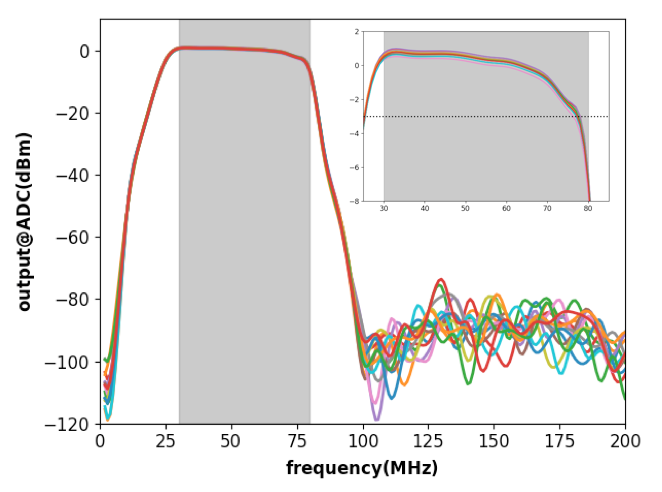}
\caption{Bandpass filter response in the RCU, $G_{\mathrm{RCU}}(\nu)$.  Different lines represent different receiving units.  The inset shows a detailed view of the filter response between $30-80$ MHz, and emphasizes the rapid fall off of the passband response at higher frequencies. The dashed line in the inset represents the 3 dB point, which occurs at 78 MHz.}
\label{fig:rcu_passband}
\end{figure}

The last stage of the signal chain is the RCU, where the signal is amplified and digitized.  Here, there are noise contributions from the amplifier, $N_{\mathrm{RCU}}$, as well as contributions from quantization and jitter generated in the analogue-digital conversion, $N_{\mathrm{ADC}}$.  The bandpass filter response in the RCU, $G_{\mathrm{RCU}}(\nu)$, has been measured, and is shown in Figure \ref{fig:rcu_passband}.  Each unit has a slightly different response, represented by the different colored lines.  We use the average response.  Any uncertainties introduced by using the average will be folded into the systematic uncertainty on electronic noise, discussed in Section \ref{sec:results}.  The 3 dB point occurs at 78 MHz.  A scale factor, $S$ is included to account for the conversion to ADC units and the overall amplification of the signal.  The total simulated power is then

\begin{equation}\label{eq:full_chain}
\begin{split}
    P_{\mathrm{sim,3}}(t,\nu)=\bigg[\bigg( P_{\mathrm{sky}}(t,\nu) + N_{\mathrm{LNA}}\bigg) G_{\mathrm{ant}}(\nu) A(\nu)L_{\mathrm{coax}}(\nu) \\+N_{\mathrm{RCU}}\bigg]G_{\mathrm{RCU}}(\nu)S+N_{\mathrm{ADC}}.
\end{split}
\end{equation}

\section{Results}\label{sec:results}

The model represented by equation \ref{eq:full_chain} includes four constant, unknown values: $N_{\mathrm{LNA}}$, $N_{\mathrm{RCU}}$, $N_{\mathrm{ADC}}$, and $S$.  The frequency dependent parts of the signal chain, $G_{\mathrm{ant}}(\nu)$, $L_{\mathrm{coax}}(\nu)$, and $G_{\mathrm{RCU}}(\nu)$, come from simulations, data sheets, and measurements, respectively.  The antenna correction factor, $A(\nu)$, is the only remaining unknown and must be solved for in a fitting procedure.  The constant unknown values $N_{\mathrm{LNA}}$, $N_{\mathrm{RCU}}$, $N_{\mathrm{ADC}}$, and $S$ are found simultaneously by minimizing the difference between $P_{\mathrm{sim}}(t,\nu)$ and $P_{\mathrm{m}}(t,\nu)$ for all 15-minute LST intervals and 1 MHz frequency bins using a least-squares fit, and are discussed further in Section \ref{sec:fit}.  Once the constants are known, the calibration factor can be found.  The calibration factor and uncertainties are discussed in Section~\ref{sec:cal}.

\subsection{Electronic Noise Contributions}\label{sec:fit}

\begin{figure}[h!]
\centering
\includegraphics[scale=0.50]{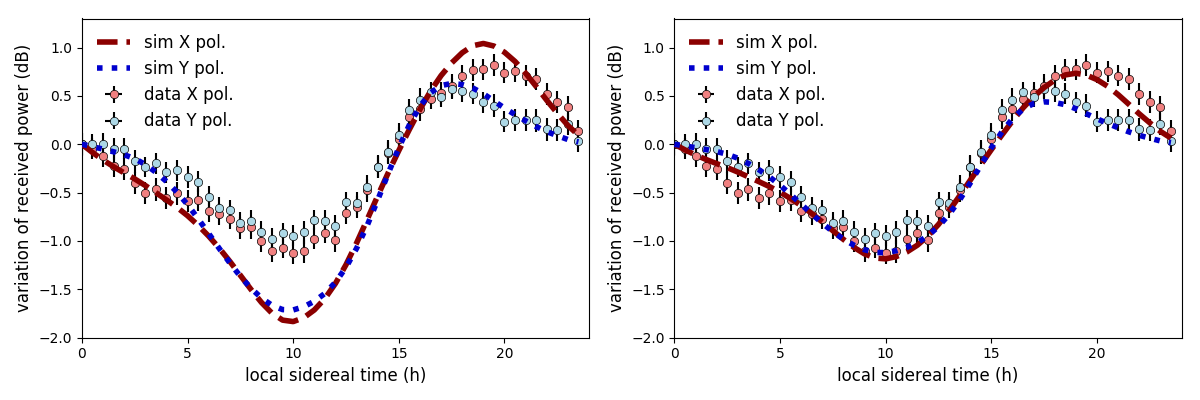}
\caption{Left: Variation of received power as a function of local sidereal time.  Data points represent the average measured power received over all antennas.  Simulated power for X and Y dipoles are represented with dark red and blue dashed lines, respectively.  Right: Simulations including thermal noise contributions, and propagated through the signal chain.}
\label{fig:LST_variation}
\end{figure}

The results of including electronic noise when comparing the power from Galactic emission to data are presented in Figure~\ref{fig:LST_variation}.  The variation in received power, referenced to LST=0, is shown as a function of LST, for both LBA dipoles.  The measured power is the same in both panels, and represents the average power received over all antennas in the Superterp stations.  In the left panel, the variation of the Galactic power without electronic noise included is shown.  It is clear that the variation in power due to the Galaxy alone is larger than for the measured data, which indicates that a component of the model is missing.  On the right, time-independent noise is included, increasing the total received power.  This decreases the variation in time, and the agreement between expectation and measurements improves.

\begin{figure}[h!]
\centering
\includegraphics[scale=0.40,trim={0cm 3cm 0cm 4cm}]{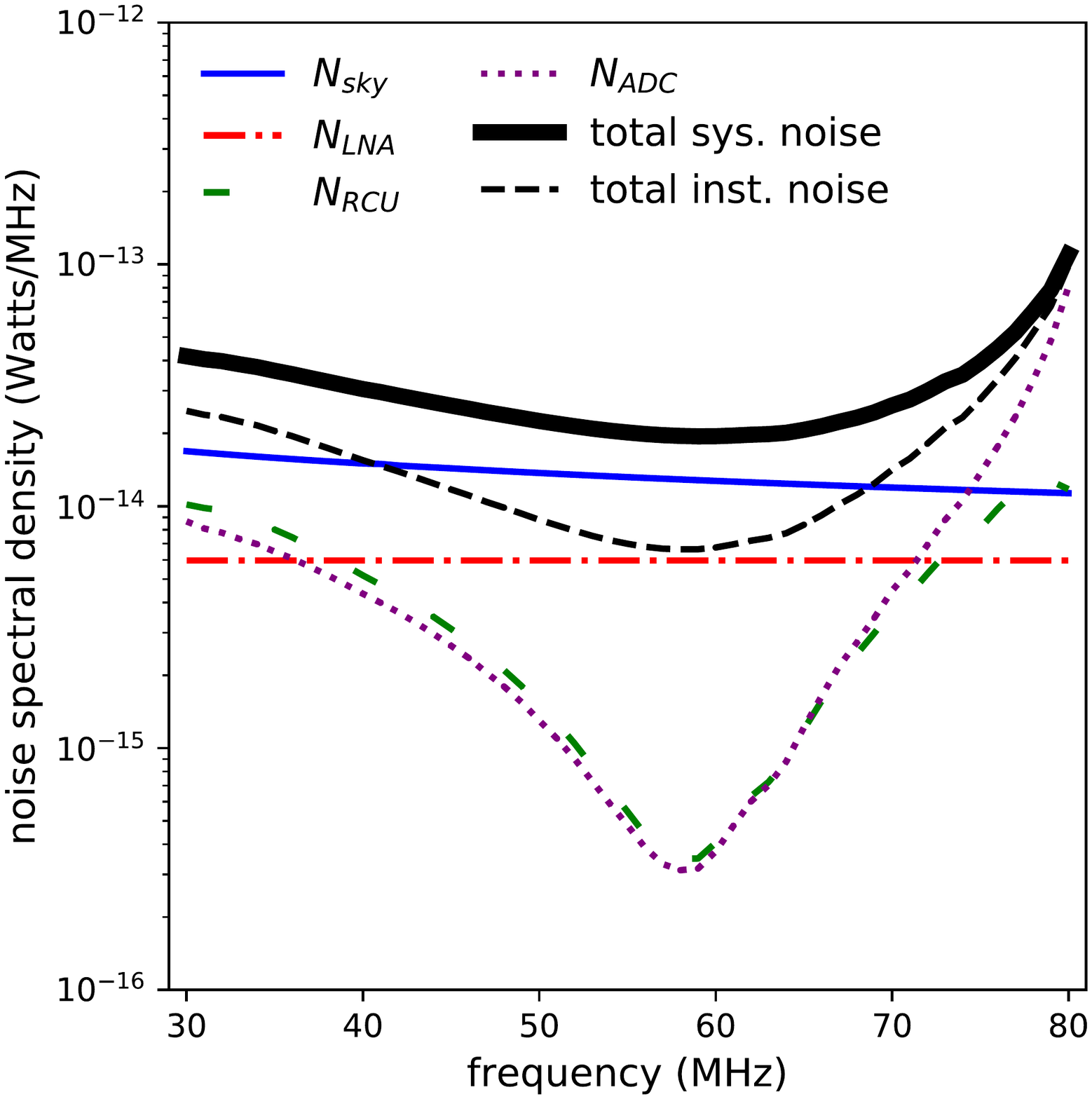}
\caption{Noise referenced to the beginning of the signal chain in units of watts/MHz.  The noise values are constant at the point they enter the signal chain, but gain a frequency dependence when propagated to the beginning of the signal chain via equation \ref{eq:full_chain}.  The sky power has been convolved with the directional response of the antenna.  Total system noise refers to the total power in the system, including contributions from both electronics and the sky.  Total instrumental noise only includes signal chain electronic noise contributions.}
\label{noise_adc}
\end{figure}

The best fit noise values are shown in Figure~\ref{noise_adc}.  Since each noise contribution comes from a difference place in the signal chain, they have much different values due to the various gains and losses, and so it doesn't make sense to compare them directly.  Instead, each value is referenced to the beginning of the signal chain, using equation \ref{eq:full_chain}.  Although each value is constant at the point it is added, propagating the values through the signal chain introduces a frequency dependence.  The sky temperature has been convolved with the directional component of the  antenna model, $D(\theta,\phi)$.  The signal from the sky dominates between 40 and 65 MHz.  Due to the antenna response, fall off of Galactic emission at high frequencies, and the RCU bandpass response, instrumental noise dominates at the highest and lowest frequencies.

\subsection{Calibration Factor}\label{sec:cal}

Once the noise values have been found, we solve for the calibration factor as
\begin{equation}\label{cal_equation}
    C^2(\nu)=\large[A(\nu)L_{coax}(\nu)G_{RCU}(\nu)S\large]^{-1}.
\end{equation}
The antenna gain is not included in the calibration factor.  Measured cosmic-ray signals originate from different directions, and so the antenna response must be considered on an event-to-event basis.  The noise values are also not included in the calibration factor, but are necessary to find the correct values for $A(\nu)$ and $S$.  For a strong cosmic-ray pulse, the power from the signal will be well above the noise.  For signals close to the noise floor, the noise, particularly $N_{ADC}$, can affect especially the highest frequencies, and can be subtracted if necessary.

\begin{figure}[h!]
\centering
\includegraphics[scale=0.8]{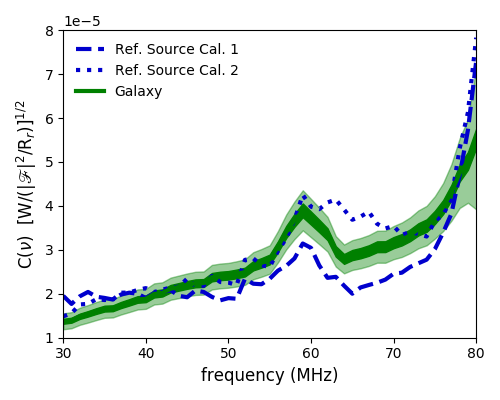}
\caption{Calibration factor for LOFAR data as determined using Galactic emission and electronic noise (in green), and compared to the calibration factors derived using the reference source described in Section~\ref{sec:intro} (blue).  The different reference source calibrations come from different calibrations of the reference source antenna itself~\cite{comm2}.  This is discussed in \ref{sec:ref_source}.  The width of the dark green line represents statistical uncertainties.  The systematics are indicated by the light green band.}
\label{calibration_curves}
\end{figure}

The resulting calibration factor is shown in Figure~\ref{calibration_curves}, indicated by the green band.  The width of the dark green band indicates statistical uncertainties, and the light green band indicates systematic uncertainties.  Two calibration factors derived from the reference source method described in~\cite{Nelles:2015gca} and \ref{sec:ref_source} are also shown in Figure~\ref{calibration_curves}.  As discussed in Section~\ref{sec:intro}, the manufacturer of the reference source provided two different characterizations of the reference source itself.  The frequency response of the reference source was not consistent between the two methods, and the differences propagate into the calibration factor of the LBAs.  This is especially evident in the resulting calibration factors between $55-75$ MHz.

The systematic uncertainties shown on the Galactic calibration are summarized in Table~1.  They are dominated by the predictions of the LFmap brightness temperature model.  Uncertainties on the absolute scaling of the brightness temperature are inherited from the sky maps used to generate $T_{\mathrm{Gal}}(\nu,\alpha,\delta)$, and are conservatively estimated at 20\% in power.  The uncertainty on how well LFmap describes the underlying maps is estimated at 10\% in power.  Finally, a 5\% uncertainty in power is included to take into account the differences between LFmap and different sky brightness temperature modeling packages~\cite{karskensThesis}.  When propagated through the calibration process, these uncertainties contribute an 11\% uncertainty to the calibration factor.  The systematic uncertainties in the brightness temperature model are discussed in more detail in \ref{sec:gal_modeling}.

We also make an estimate of systematic uncertainties introduced by the WIPL-D simulated antenna pattern.  From the previous calibration campaign using a reference antenna attached to an octocopter, a comparison between the measured and predicted gain patterns of the LBAs showed reasonable agreement \cite{Nelles:2015gca,krauseThesis}. In order to estimate the quantitative uncertainty on $D(\theta,\phi)$, we adjust the gain pattern of the LBAs so that power received at $\theta=0$ varies by $\pm 10\%$, while conserving the total power received over all zenith angles.  We found that if the power received at zenith varies by more than $\pm 10\%$, or if the shape of the gain pattern is fundamentally altered, the LST variations shown in Figure \ref{fig:LST_variation} cannot be reproduced.  For this reason, we have chosen $\pm 10\%$ as a conservative estimate for fluctuation in gain pattern.  Once propagated through the calibration procedure, this has the effect of changing the final calibration factor by a maximum of $2.5\%$, which we take as an estimate of the uncertainty in the directionality of the antenna model as it applies to the calibration.  For cosmic-ray events, the antenna response is not integrated over the whole sky, but instead the response at a particular direction on the sky must be used.  From the octocopter campaign measurements we find that at 50 and 60~MHz there is negligible systematic difference between measured data and predictions based on the WIPL-D simulation as a function of zenith angle~\cite{krauseThesis}.  This is the frequency band in which the antenna has the largest gain, and is what dominates the LOFAR X$_\textrm{max}$ and energy scale analyses.

The uncertainty in the electronic noise contribution comprises the remaining systematic uncertainties.  The noise values found from the fitting procedure represent averages over all antennas.  In practice, each antenna has slightly different noise values and a different RCU response, resulting in slightly different calibration factors.  The electronic noise uncertainty is found by using the fitted noise values to derive individual calibrations for each antenna.  The spread of these curves is frequency dependent, and represents the systematic uncertainty in the electronic noise.  Below 77 MHz, this contribution to the systematics is at most 6.5\%.  Above 77 MHz, this increases to $\sim$ 20\% due to instabilities introduced by the rapid fall off of the RCU bandpass filter.  

Statistical uncertainties are dominated by event-to-event fluctuations in the background power of each antenna, which are on the order of 5\%, and presumably due to environmental factors.  The fluctuations between antennas for the same event are of the same order, but this uncertainty is folded into the uncertainty in electronic noise.

The features of the new Galaxy calibration can be explained as follows.  The overall rise in calibration factor with frequency is due to the frequency dependent loss in the cables.  The more pronounced rise at the highest frequencies compensates for the bandpass filter in the RCU.  The feature close to 58 MHz is due to a misalignment of the resonance frequency in the antenna model simulations, which occurs because the position of the resonance frequency changes in different environmental conditions.  This feature enters the calibration factor via the $A(\nu)$ correction factor. 

The Galaxy calibration method has the advantage of being easily repeatable.  For example, the same procedure can be repeated in different time intervals to determine if the calibration factor changes over time.  Between 2012 and 2018, the LBA calibration fluctuates on the order of 5\% between each year long period, but does not show any systematic trend over time.  Additionally, with enough background data, calibration factors can be found for individual antennas.  At the moment, we use one calibration for all antennas and dipoles.

\begin{table}
\begin{center}\label{cal_uncertainties}
 \begin{tabular}{c c} 
 
 \textbf{Systematic Uncertainty} & \textbf{Percentage} \\ [0.5ex] 
 \hline\hline
 antenna model & 2.5\\
 \hline
 sky model & 11\\
 \hline
 electronic noise $<$ 77 MHz & 6.5\\
 \hline
 electronic noise $>$ 77 MHz & 20\\

 \hline
 \hline
 \textbf{total $<$ 77 MHz} & \textbf{13}

\end{tabular}
\end{center}
\caption{Summary of the systematic uncertainties in the LBA Galactic calibration.}
\end{table}

\section{Discussion}\label{sec:discussion}

The purpose of this work is to provide an absolute, frequency-dependent calibration for the LBAs with so as to analyze radio pulses from cosmic rays.  In Section~\ref{sec:compare_MC}, we present results of applying the new calibration to cosmic-ray data, and make a comparison with the reference source calibrations and CoREAS simulations~\cite{Huege:2013vt}.  In Section~\ref{sec:compare_TR}, using results from the Tunka-Rex collaboration, we demonstrate that there are systematic trends in the reference source over the $30-80$ MHz band that are evident both in Tunka-Rex analyses and in the LOFAR calibration factors.

\subsection{Comparison with air shower simulations}\label{sec:compare_MC}

The frequency spectra of cosmic-ray radio pulses are generally expected to be of exponential form in the $30-80$~MHz band~\cite{Huege:2010vm}.  This behavior is seen in CoREAS simulations~\cite{Huege:2013vt}.  The spectrum is expected to be smooth and without any kinks in the shape.  This is an important feature, as the spectral index changes with distance to the shower core, and also has a dependence on shower development~\cite{rossettoICRC2017,jansenThesis}.  In order to compare calibrated LOFAR data with air shower simulations, we check that the slopes of the logarithm of the power spectra are in agreement, and that the slope doesn't change over the frequency range of $30-80$ MHz.

The data set for this comparison consists of the 20 strongest cosmic-ray events.  For each shower, measurements from 48 antennas from several stations are used, giving a total of 1653 individual signals.  The events were processed in the standard pipeline~\cite{schellart2013}.  We look at the total power spectra of the reconstructed electric fields at the position of the antenna.  The average background is subtracted from the cosmic-ray signal.  We compare data calibrated with the new Galaxy method and both existing reference source methods.  CoREAS simulations were run for each antenna for each event because frequency spectra depend on the observer positions.  Simulations used input parameters such as energy and shower geometry based on previous LOFAR analyses~\cite{buitinkNature2016}.

\begin{figure}[h!]
\centering
\includegraphics[scale=0.58,trim={2cm 2cm 5cm 1cm},angle =90]{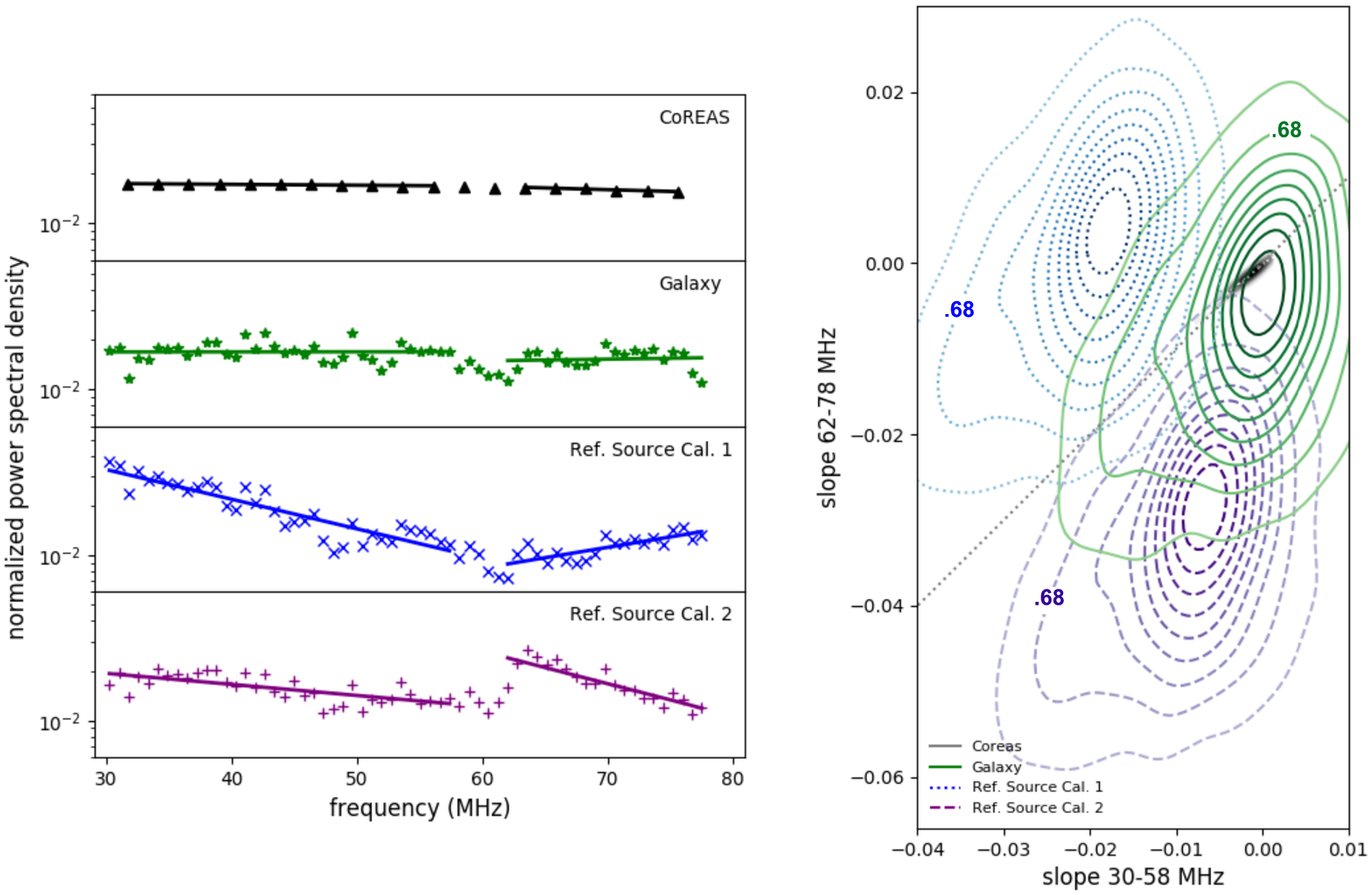}
\caption{Results of comparing calibrated LOFAR data with CoREAS simulations.  Left:  Fits to the power spectrum of a typical event, comparing different calibration methods and simulation.  The points represent the measured power spectra, and the lines represent linear fits in the ranges $30-58$~MHz and $62-78$~MHz.  The purple and blue points indicate the data calibrated using the two reference source methods, and the green points indicate data calibrated with the Galactic method.  Black points indicate simulation.  Right:~Contour plot showing the relation between the slopes before and after the LBA resonance frequency.  The 68\% confidence interval is indicated for each set of calibrated data.}
\label{power_spec_ex}
\end{figure}

A straight line was fit to each power spectrum in the $30-58$ MHz and $62-78$ MHz bands.  The highest frequencies have been excluded due to the increased uncertainty in the electronic noise characterization.  The frequencies close to the resonance peak were also excluded as the resonance peak can differ slightly in different circumstances, and so can be unstable.  A typical example of power spectra and  linear fits is shown in the left panel of Figure~\ref{power_spec_ex}.  The CoREAS signal can be fit well with a straight line across the full frequency band.  It is also apparent that in the cases of the signals calibrated with the reference source, the shape of the spectra are inconsistent over the band, while the Galaxy calibrated data is more consistent.

In order to look at trends over the whole data set, a contour plot relating the slopes between $30-58$ MHz to the slopes between $62-78$ is shown in the right panel of Figure~\ref{power_spec_ex}.  The 68\% confidence interval is marked for each calibration, indicating the contour line containing points that lie within $\pm$ 1 standard deviation  of the mean.  The points representing  CoREAS simulations fall on the diagonal line, showing that the slope is consistent over both low and high bands.  The revised Galaxy calibration shows promising agreement with simulation, both in slope amplitude, and in consistency across the entire band.

\subsection{Comparison with Tunka-Rex}\label{sec:compare_TR}

Other radio cosmic-ray experiments, including LOPES~\cite{Nehls:2008ix,LOPES:2015eya} and Tunka-Rex~\cite{Bezyazeekov:2015rpa} used the same reference source for calibration, and so the questions regarding the calibration of the reference source antenna are relevant for them as well.  One way to compare results between experiments is to look for systematic behavior as a function of frequency.  In this analysis, we have used the same reference source calibration as Tunka-Rex, corresponding to ``Ref. Source Cal. 1" in Figure \ref{calibration_curves}.

\begin{figure}[h!]
\centering
\includegraphics[scale=0.5]{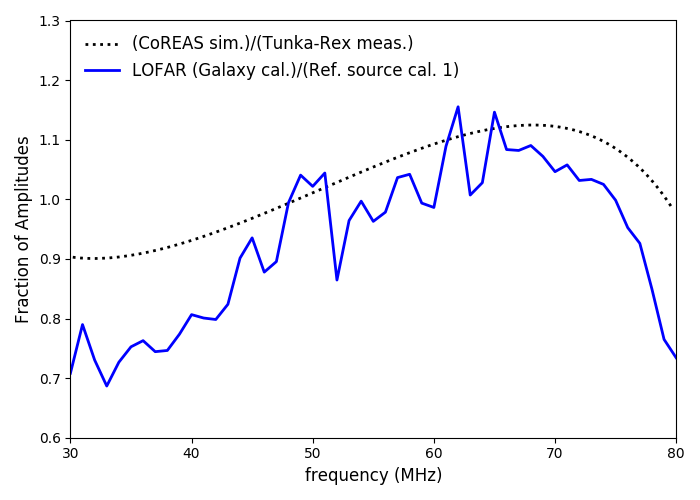}
\caption{Frequency dependent systematics for Tunka-Rex simulations and data, and LOFAR calibrations.  The dashed line indicates the average ratio of the amplitude of CoREAS simulations to measured Tunka-Rex data. The solid line is the ratio of LOFAR Galactic calibration to the Ref. Source Cal. 1.}
\label{tunka_compare}
\end{figure}

To study frequency dependent systematics of Tunka-Rex data, we present analysis done by the Tunka-Rex collaboration \cite{kostunin,comm1}.  The amplitude of measured cosmic-ray signals is compared to the amplitude of signals simulated with CoREAS.  Simulations were run taking into account specific event geometry, energy, and $X_{\textrm{max}}$.  The average ratio of simulated to measured amplitude for many air showers and antennas is shown as a dashed line in Figure \ref{tunka_compare}.  The change in the ratio over the frequency band indicates that there is discrepancy between simulated pulse shapes and the pulse shapes of the Tunka-Rex data, which depend on the calibration derived from the reference source antenna.  We note that the size and frequency dependence of the observed effect is within the uncertainties and taken into account in previous Tunka-Rex analyses.

To study systematic effects in LOFAR data, we look at the ratio of Galaxy calibration values to reference source calibration values. The same frequency dependent trend is seen over the $30-80$~MHz band as is seen in the Tunka-Rex analysis, and the fraction of amplitudes is consistent within systematic uncertainties.  This indicates that there are systematic problems with the reference source between $30-80$ MHz.  Because the existing calibrations suffer from the same problem,  adopting this new calibration technique will increase the consistency between measurement and Monte Carlo, both for LOFAR and Tunka-Rex data.

\section{Conclusions}

A frequency dependent antenna calibration with low systematic uncertainties is critical for studies of cosmic-ray radio data.  In order to calibrate the LOFAR LBAs, a reference signal is required.  Previously, the LBAs were calibrated using two different methods.  One made use of an externally calibrated reference source.  The other used a combination of Galactic emission and electronic noise as a reference signal, but did not model the signal chain.  Both methods agreed in overall amplitude of the calibrated signal, but produced differences in the frequency response.  In particular, there is conflicting information about the frequency dependence of the reference source, which directly propagates into the LOFAR calibration.  The systematic uncertainty in the electronic noise estimates for the Galactic emission method limited the usefulness of that calibration.

In order to proceed with spectral analyses, the calibration of the LBAs has been revisited, using Galactic emission and a detailed model of the LOFAR signal chain as a calibration source with which to compare measured data.  Each step of the signal chain is now independently modeled, and electronic noise values are added where noise is expected to enter the system.  Modeling the noise in this way decreases the systematic uncertainties and accounts for the frequency dependence of each signal chain component.  The systematic uncertainties in the calibration are now limited by the reference measurements of the Galactic background.  

To compare LOFAR data calibrated with the new Galactic method to air shower simulations, we looked at the slope of the frequency spectra of observed and simulated signals.  The behavior of the spectral slope for LOFAR data using the new calibration shows promising agreement with CoREAS simulations, both in amplitude, and consistency over the entire $30-80$ MHz band.  Furthermore, the frequency dependent systematic effects seen between the new Galactic calibration and old reference source calibration show the same trends as systematic effects between CoREAS simulations and Tunka-Rex data, which were calibrated using the same reference source.  With the new Galactic calibration, detailed frequency analyses are now possible.  This method has the benefit of being easily repeatable, and applicable to any radio experiment with a view of the sky and knowledge of the signal chain.

 \section*{Acknowledgements}

The authors would like to thank Frank Schr\"{o}der and Dmitriy Kostunin for their help comparing LOFAR results with Tunka-Rex results, and for the preparation of Figure \ref{tunka_compare}.   The LOFAR cosmic-ray key science project acknowledges funding from an Advanced Grant of the European Research Council (FP/2007-2013) / ERC Grant Agreement n. 227610. The project has also received funding from the European Research Council (ERC) under the European Union’s Horizon 2020 research and innovation programme (grant agreement No 640130). We furthermore acknowledge financial support from FOM, (FOM-project 12PR304). TW is supported by DFG grant WI 4946/1-1. AN is supported by the DFG grant NE 2031/2-1. LOFAR, the Low Frequency Array designed and constructed by ASTRON, has facilities in several countries, that are owned by various parties (each with their own funding sources), and that are collectively operated by the International LOFAR Telescope foundation under a joint scientific policy.

\newpage

\appendix

\section{Reference Source Calibration}
\label{sec:ref_source}
\begin{figure}[h!]
\centering
\includegraphics[scale=0.49]{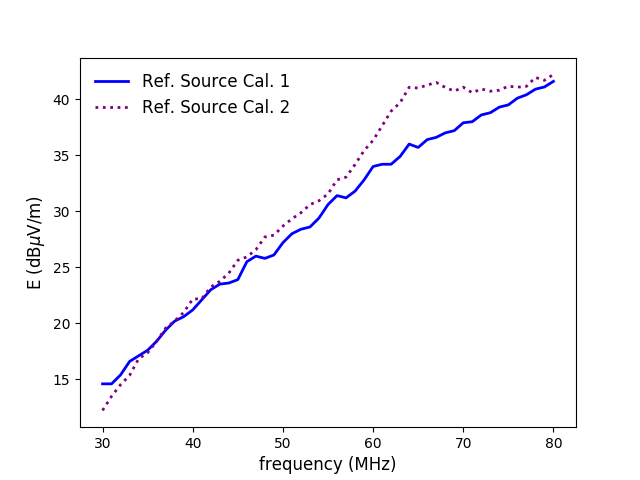}
\caption{Electric field strength of the radiation source as a function of frequency.  The dashed line indicates the results calibrated in  a GTEM cell, and the solid line indicates results from calibration in an anechoic chamber.  The field strength of the reference source antenna (VSQ 1000) is scaled to 10m free space conditions.} 
\label{ref_source_efield}
\end{figure}

This appendix includes a brief summary of the reference source calibration method.  As for the Galaxy method, the calibration factor $C(\nu)$ for the reference source method is defined as 
\begin{equation} \label{eq:cal_factor}
 C^2(\nu) = \frac{P_{e}(\nu)}{P_{m}(\nu)}.
\end{equation}
Here,~$P_{e}(\nu)$ is the expected power induced in the antenna from the reference source, and $P_{m}(\nu)$~is the corresponding power measured by the LBAs.  Data for this calibration were collected during the calibration campaign performed on May 2014 using a VSQ 1000 reference radiation source, consisting of a DPA 4000 biconal antenna and RSG 1000 signal generator~\cite{Nelles:2015gca,teseq}. Since the LOFAR calibration relies on knowledge of the power emitted by the reference source, the characteristics of the reference source itself must be well known.    This reference source was originally characterized between $30-1000$~MHz in a Standard 3~m Anechoic Chamber~(SAC) at 3~m distance and 1.5~m height with 10 MHz resolution.   We now have specifications of the reference source between $30-100$~MHz in 1 MHz resolution, with measurements made both in a SAC and a Gigahertz Transverse Electromagnetic cell~(GTEM) at 3~m distance \cite{comm2}.  Figure~\ref{ref_source_efield} shows the strength of the emitted electric field in the frequency range 30$-$80~MHz for the two characterization procedures. The major difference between the two characterization methods is that the anechoich chamber has a reflective floor, while the GTEM cell does not. Thus, while frequency-dependent reflections take place during measurements performed in the anechoic chamber, they are not included in measurements performed in the GTEM cell.  This affects the characterized electric field of the reference source. Additionally, neither measurement was made in the far-field of the reference antenna, although both results were re-scaled to a 10~m free space condition.

\section{Antenna Theory}
\label{sec:cal_theory}

\begin{figure}[h!]
\centering
\includegraphics[scale=0.7,trim={5cm 0cm 5cm 0.5cm}]{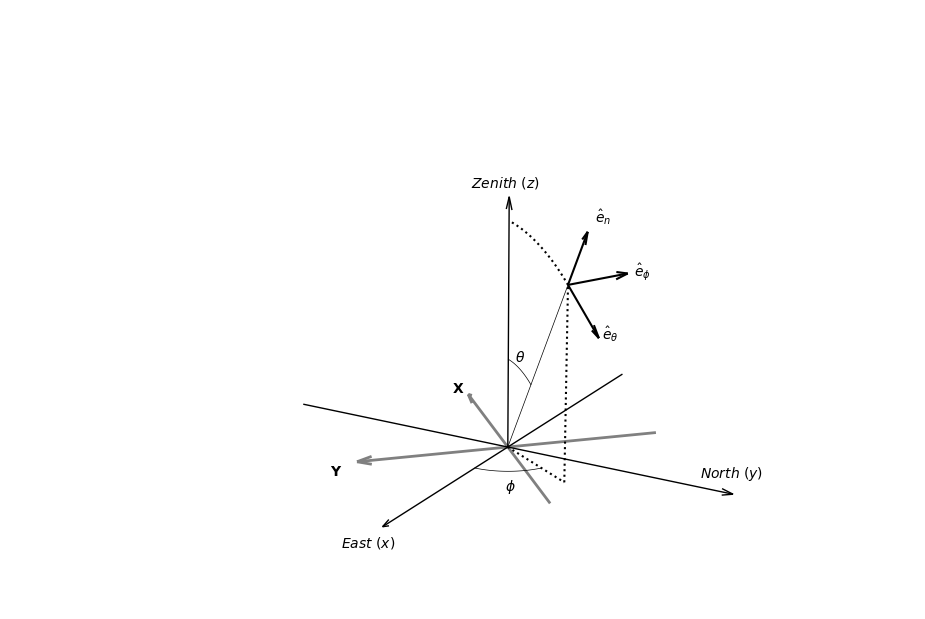}
\caption{Coordinate system used for LOFAR cosmic-ray analysis, where angles $\theta$ and $\phi$ denote the zenith and azimuth angles, respectively.  The X and Y arrows denote the LBA dipoles.}
\label{coordinates_schematic}
\end{figure}

The following derivation relates the effective height (or antenna response) of the antenna, $\mathbf{H}(\nu,\theta,\phi)$, the effective area of an antenna, $A(\nu,\theta,\phi)$.  The coordinate system used is shown in Figure \ref{coordinates_schematic}, where  the X and Y arrows denote the X and Y polarizations of the LBAs.

A complex electric field $\mathbf{E}(\nu)$ in the frequency domain arriving from the $(\theta,\phi)$ direction can be decomposed into two components perpendicular to the direction of propagation as 
\begin{equation}
    \mathbf{E}(\nu,\theta,\phi)= E_{\theta}(\nu,\theta,\phi)\mathbf{\hat{e}_{\theta}}+ E_{\phi}(\nu,\theta,\phi)\mathbf{\hat{e}_{\phi}}.
\end{equation}
The instantaneous Poynting vector, or power density, of the incoming wave in the frequency domain can be written
\begin{equation}
    S(\nu,\theta,\phi)=\frac{[E^2_\theta(\nu,\theta,\phi)+E^2_\phi(\nu,\theta,\phi)]}{Z_0} 
\end{equation}
where $E_\theta, E_\phi$ are the $\theta,\phi$ components of the electric field and $Z_0$ is the impedance of free space \cite{kraus1988}.  The power an antenna receives can be written
\begin{equation}\label{eq:PSA}
    P(\nu,\theta,\phi)=S(\nu,\theta,\phi)A(\nu,\theta,\phi). 
\end{equation}
This relation is used in Section \ref{sec:cal_gal} to derive the sky power used in the Galaxy calibration.  We can write the power received by the antenna as 
\begin{equation}\label{p_sa}
    P(\nu,\theta,\phi)=S(\nu,\theta,\phi)A(\nu,\theta,\phi)=\frac{E^2(\nu,\theta,\phi)A_e(\nu,\theta,\phi)}{Z_0}.
\end{equation}

Another relation is the vector effective length, $\mathbf{H}(\nu,\theta,\phi)$, which relates the electric field at an antenna to the voltage induced in the antenna, as  
\begin{equation}
    V(\nu,\theta,\phi)=\mathbf{H}(\nu,\theta,\phi) \cdot \mathbf{E}(\nu,\theta,\phi).
\end{equation}
For the LOFAR dipole antennas, the vector effective length can be expressed using the Jones Matrix.  For a given frequency, the voltage in each dipole polarization can be written

\begin{equation}
    \begin{pmatrix}
    V_{X}(\nu)\\
    V_{Y}(\nu)
    \end{pmatrix}
=
    \begin{pmatrix}
    J_{X\theta}(\nu) & J_{X\phi}(\nu)\\
    J_{Y\theta}(\nu) & J_{Y\phi}(\nu)
    \end{pmatrix}
    \begin{pmatrix}
    E_{\theta}(\nu)\\
    E_{\phi}(\nu)
    \end{pmatrix}.
\end{equation}

For an antenna with radiation resistances matched to the load, we can also write the power received by the antenna as
\begin{equation}\label{eq:P_hE}
    P(\nu,\theta,\phi)=\frac{V^2(\nu,\theta,\phi)}{R_r} = \frac{H^2(\nu,\theta,\phi)E^2(\nu,\theta,\phi)}{R_r}.
\end{equation}
Relating equations \ref{p_sa} and \ref{eq:P_hE}, we arrive at the relation between effective height and effective area,
\begin{equation}\label{eq:AtoH}
    A(\nu,\theta,\phi)=\frac{H^2(\nu,\theta,\phi) Z_0}{R_r}.
\end{equation}

\section{Systematic Uncertainties of the Sky Brightness Temperature}
\label{sec:gal_modeling}
The largest systematic uncertainty on the Galactic calibration procedure comes from modeling the sky brightness temperature.  A number of different models exist that interpolate sky maps at specific frequencies to predict the brightness temperature at any position on the sky and at any frequency in the range of tens of MHz to a few GHz.  Two models in particular, LFmap~\cite{LFmap} and Global Sky Model (GSM)~\cite{galactic_radio_emission_map} were compared, and it was determined that the average sky brightness temperature in the band 30--80~MHz differed by at most 5\%~\cite{karskensThesis}.  In this work LFmap is used to predict the sky temperature~\cite{LFmap}.  LFmap was originally developed for the use of the Long Wavelength Array (LWA)~\cite{Ellingson:2012rc,Dowell:2012rt}, and models the sky brightness temperature as a combination of cosmic microwave background, isotropic emission, and Galactic emission.  The temperature at a given frequency $\nu$, right ascension $\alpha$, and declination $\delta$, is given as
\begin{equation}
    T_{\mathrm{sky}}(\nu,\alpha,\delta)=T_{\mathrm{CMB}}+T_{\mathrm{Iso}}(\nu)+T_{\mathrm{Gal}}(\nu,\alpha,\delta).
\end{equation}
The $T_{CMB}$ contribution is from the cosmic microwave background and is a constant $2.73$~K.  The isotropic contribution, $T_{Iso}(\nu)$, is mainly attributed to unresolved emission from extragalactic sources and follows the work of Lawson et al.~\cite{lawson1987}, and~ Bridle et al.~\cite{bridle1967} cited therein, where the emission is characterized as
\begin{equation}
T_{Iso}(\nu)=50\textrm{K}\big(\frac{150 \textrm{ MHz}}{\nu}\big)^{2.75}.
\end{equation}
The exponent 2.75 has an uncertainty of $\pm 0.2$.  Since the isotropic contribution contributes at maximum 20\% to the total brightness temperature, the uncertainty in the isotropic component propagates to 5\% of the total brightness temperature.

The Galactic contribution is due to synchrotron radiation from electrons in the Galactic magnetic field~\cite{rybicki1979radiative}, and can be found at each point on the sky by using the temperature measured at frequency $\nu_0$ and position $(\alpha,\delta)$.  Assuming the temperature follows a power law with spectral index $\beta$,  the Galactic contribution to brightness temperature is
\begin{equation}
    T_{Gal}(\nu,\alpha,\delta)=T_{Gal}(\nu_0,\alpha,\delta)\big(\frac{\nu_0}{\nu}\big)^{\beta}.
\end{equation}
At 408~MHz, a sky map has been produced by Haslam et al.~\cite{Haslam:1982zz}.  Platania et al. cleaned the Haslam map for artifacts, combined it with maps at 1420~MHz and 2326~MHz, and derived the spectral indices for Galactic emission~\cite{Platania:2003ta}.  This approach does not account for spectral bending below 200~MHz that results from the flattening of the Galactic electron spectrum below 3~GeV, or HII absorption regions below 45~MHz.  LFmap corrects for this using the 22~MHz map from Roger et al.~\cite{Roger:1999jy}. The uncertainty on the absolute calibration of these underlying maps propagates into the LFmap model.  The uncertainty in the 408~MHz map is quoted as 10\%, and the 22~MHz map 16\%.   The higher frequency maps have lower uncertainties.  The uncertainty of the absolute scaling of the model is therefore estimated to be 20\%.  This level of uncertainty is consistent with the underlying maps used in other modeling procedures, such as GSM.

We also estimate how well the model describes the underlying maps.  The LFmap documentation does not provide a value for this.  Since the average sky temperature predicted by LFmap and GSM are consistent to within 5\%, we use the uncertainty quoted for GSM.  The GSM approach is different than that of LFmap.  GSM uses a Principle Component Analysis (PCA)~\cite{Jolliffe:1986}
and does a 3 component fit over 11 sky maps between 0.010 and 94~GHz to predict a sky brightness temperature.  The uncertainty in how well the model represents the underlying maps is determined by repeating the fitting procedure, using a fit done with 10 maps to predict the remaining map.  The resulting uncertainty is conservatively estimated to be 10\%.

In total, we combine the 5\% choice of model uncertainty, the 20\% uncertainty in the absolute scaling of the underlying maps, and the 10\% uncertainty in fitting procedure to reach a total of 23\% uncertainty in the sky temperature.  The propagates into an uncertainty of 11\% in the calibration value.



\bibliographystyle{elsarticle-num}

\bibliography{main}

\end{document}